\documentclass[prd,aps,epsf,bm,floats,amsfonts,amssymb,amsmath,nobibnotes,nofootinbib,showkeys]{revtex4}
\begin{document}

\def\be{\begin{equation}}
\def\ee{\end{equation}}
\def\al{\alpha}
\def\bea{\begin{eqnarray}}
\def\eea{\end{eqnarray}}

\title{A comprehensive view of  cosmological Dark Side}%
\author{S. Capozziello$^{a, b}$ and G. Lambiase$^{c, d}$} 
\affiliation{$^a$Dipartimento di Fisica, Universit\'a di Napoli "Federico II", Via Cinthia, I-80126 - Napoli, Italy.}
\affiliation{$^b$INFN Sez. di Napoli, Compl. Univ. di Monte S. Angelo, Edificio G, Via Cinthia, I-80126 - Napoli, Italy.}
\affiliation{$^c$Dipartimento di Fisica "E.R. Caianiello" Universit\'a di Salerno, I-84081 Lancusi (Sa), Italy,}
\affiliation{$^d$INFN - Gruppo Collegato di Salerno, Italy.}

\date{\today}

\begin{abstract}
Recent cosmological and astrophysical observations point out  that the Universe is in accelerating expansion and filled up with non-luminous matter.
In order to explain  the observed large scale structures and this  accelerating behavior one needs  a huge amounts of Dark Energy and Dark Matter. Although many attempts have been done, both at theoretical and experimental level, up to now there are several  models proposed to explain such mysterious components. However, no final conclusion on the nature of dark components has been reached up to now so the question is completely open.  In this paper, we review, with no claim of completeness,   the recent results and ideas underlying Dark Energy and Dark Matter issues
\end{abstract}


\maketitle

\section{Introduction}

One of the greatest challenges in physics is nowadays to understand the nature of
Dark Matter (DM)  as well as of  Dark Energy (DE). Both issues are fundamental
in particle physics and cosmology, and, although many ideas and proposals have been developed,
till now no fully satisfactory theory or model has been found and DM and DE escape any final explanation at fundamental level.

Specifically DM is a  non-luminous and non-absorbing form of matter  interacting
only gravitationally. For this reason,  it is very difficult to achieve a precise
indication on its nature at fundamental level. However, it
 plays a relevant role for the formation of the large scale structure,
and should be related to some new unknown particle. DE,
an unknown (unclustered) form of energy, is instead responsible for the observed acceleration
of the present Universe. Recent observations point out it should constitute almost 75\%
of cosmic matter-energy budget, turning out that the so-called "Dark Side" problem is far to be solved.

From the experimental side, there are many observations, both at cosmological and astrophysical scales, that
support the evidence of DM: 1) Very precise measurements of  cosmic microwave
background radiation in the WMAP experiment \cite{Dunkley2009,Komatsu2011}; 2)
gravitational lensing in weak \cite{Refregier2003} and strong \cite{Tyson1998} regimes; 3) hot gas in clusters \cite{Lewis2003};
4) the Bullet Cluster \cite{Clowe2006}; 5) Big Bang nucleosynthesis \cite{FieldsSarkar2008};
6) constraints from large scale structure \cite{Allen2003};
7) distant supernovae of SNeIa type \cite{Riess1998,Perlmutter}.
Current data constrain the energy densities of the Universe in a baryonic component,
mostly known ($\Omega_B \lesssim 0.0456 \pm 0.0016$), a DM (CDM) component ($
\Omega_{\text{CDM}} \simeq 0.227 \pm 0.014$) still unknown, and a DE
component ($\Omega_\Lambda \simeq 0.728 \pm 0.015$ with a great uncertainty on the generating mechanisms).
The luminous matter in
the Universe is less than 1\% of the total composition of the Universe. The current most precise estimation of the
density of non-baryonic DM $\Omega_{\text{DM}}$ is obtained combining the measurements of the cosmic macrowave background (CMB) anisotropy and of the spatial
distribution of the galaxies and has found to be $\Omega_{\text{DM}} h^2 = 0.110 \pm 0.006$.
The "local" DM present in the Galactic disk has an average density of  $\rho^{(local)}_{\text{DM}} \simeq  0.3$GeV/cm$^3$
\cite{KamionkowskiKinkhabwala1998}.

An alternative approach to face the above issues is to consider DM and DE as the manifestation of the break-down
of General Relativity (GR) on large scales, and consider possible modifications and generalization of the
laws of gravitation \cite{capozzielloPR}.

In this review paper, we illustrate, with no claim of completeness, some aspects of DM and DE, and the main candidates to solve the
big puzzle of modern astrophysics and cosmology.  We start by recalling the basic argument of GR.
Then, in the first part ,we discuss theoretical models that try to explain the DE and the accelerating phase of the present Universe.
The second part of the review is devoted to the DM physics mainly pointing out the fundamental physics issues.

\section{Preliminary aspects}

It is well known that the best self-consistent theory of gravity that dynamically describes the space-time evolution and
matter content in the Universe is represented by General Relativity. This theory is able to explain several
gravitational phenomena, and has allowed a deepest understanding and comprehension of laws underlying the gravitational physics ranging from ordinary scales (laboratory
scales) to large scales (cosmological scales).  Predictions and confirmations of General Relativity have been mainly tested in Solar System, but as will be discussed below, maybe the cornerstone of General Relativity is represented by the Big Bang Nucleosynthesis, i.e. the formation of light elements in the Universe, and cosmic microwaves background (CMB) radiation.
However, despite all the fundamental results that General Relativity provides, it is not immune to many problems.
First of all, General Relativity presents technical and conceptual difficulties if {\it thought} as a field theory, since it is a non-renormalizable theory. Field equations of General Relativity predict the existence of gravitational waves, in analogy to Maxwell's electromagnetism where electromagnetic waves exist in the absence of sources. The quantum field associated to the gravitational field is the graviton (the analogue of the photon in Quantum Electrodynamics, QED) which is massless and carries spin s = 2. It can be considered, in the conventional approach, as a fluctuation of the geometry around the flat Minkowski space-time. According to quantum field theory, one may study processes in which the gravitons are emitted or absorbed by mass-sources, and processes of self-interaction (gravitons can be emitted and absorbed by gravitons) induced by the non-linearity of General Relativity. These processes are characterized by the dimensionless quantity $E/E_{Pl}$, where $E$ is the typical energy of the process under consideration, and $E_{Pl}\sim 10^{19}GeV$ is the Planck energy. In general,  gravitational forces manifest at regimes in which $E/E_{Pl}\ll 1$, so that the effects of quantum fluctuations are strongly suppressed and  therefore negligible. However, in the regime $E/E_{Pl} \lesssim 1$, graviton fluctuations become relevant and quantum effects play non trivial role on the physical processes. It is expected that this regime occurs in the early phase of the Universe evolution or in catastrophic collapse of matter forming black holes. In any case, it is basically the regime $E/E_{Pl} \lesssim 1$ that makes General Relativity, when treated as a standard quantum theory of the metric fluctuations around a given classical space-time, badly divergent and meaningless.
In order that a quantum field theory is well-defined, the theory must be characterized by a choice of {\it finite} parameters, which could, in principle, be set by experiment (in QED, for example, these parameters are the charge and mass of the electron, as measured at a particular energy scale). In the case of gravity, we  have not a meaningful physical theory because its quantization requires {\it infinite} parameters which are necessary to define the theory, with the consequence that one can never fix the values of every parameter by means of infinite experiments. In other words, at high energies where quantum effects becomes important, also the infinitely many unknown parameters become important, and the theory is unable to make any predictions\footnote{Although quantum mechanics and gravity are consistent and well tested at low energies, as distinct theories, problems arise when they are combined to build up a quantum theory of gravity because severe divergencies occur in computing loop diagrams of processes in which gravitons are involved. To cure these divergencies and make finite the results, one may introduce a cutoff (which is generally assumed to be of the order of the Planck scale). This option if of course far to be appealing and suggests that a new model to describe the Nature at these high energy scales is necessary. This typical issue of quantum gravity is, in principle, solved by String Theory.}
\cite{birrell,shapiro}. Clearly, at low energies, quantum gravity has to reduce, despite the unknown choices of the infinitely many parameters, to the usual General Relativity.


Let us now go back to the predictions of General Relativity. The formation of the light elements and the CMB
represent the great success of General Relativity predictions at cosmological level. Considering  the former (see \cite{garcia,kolb,bernstein,sarkar}),
it is related to the creation of  light elements towards the end of the {\it first three minutes}
which provides the deepest detailed probe for the Big Bang.
It was suggested by Eddington  (1920) (see \cite{alpher,wagoner} for a historical accounts) that the the fusion of hydrogen into helium should be the responsible of the energy of the Sun.
Hans-Bethe, ten years later, gave the detailed reactions by which stars burn hydrogen.
It was Gamow that argued that similar processes might have occurred also in the hot and dense early Universe leading to the formation
of light elements \cite{gamov}. The temperature necessary to active these processes is around
$T_{NS}\sim (0.1 \div1)$ MeV, while the density
is $\rho_{NS}=\displaystyle{\frac{\pi^2 g_*}{30}\, T_{NS}^4}\sim 82$gr/cm$^3$.
There is, however, a substantial difference regarding  the physical conditions underlying stars and Big Bang Nucleosynthesis (BBN).
In fact, although both processes are driven by identical thermonuclear reactions,
in the star formation, the gravitational collapse heats up the core of the stars and reactions last for billions of years (except in supernova explosions, which last a few minutes and creates all the heavier elements beyond Iron), while in the case of the Universe, expansion cools the hot and dense plasma in just a few minutes.
In this picture, the abundances of the light elements is correlated with
the neutron capture cross sections\footnote{Although the early period of
cosmic expansion is much shorter than the lifetime of a star, there was a large number of free neutrons
at that time, so that the lighter elements could be built up quickly by successive neutron captures, starting
with the reaction $n+p\to D+\gamma$ \cite{garcia}.}, in rough agreement with observations \cite{weinberg,burles}.
Computations in the framework of BBN are able to produce all the light elements up to Beryllium-7, while
the other nuclei (up to Iron (Fe)) are produced in heavy stars, and beyond Fe in novae and supernovae explosions \cite{olive}.
Following \cite{garcia}, only the elements $H$, ${}^4 He$, $D$, ${}^3 He$, ${}^7 Li$ (and perhaps also ${}^6 Li$) can be computed with accuracy better than 1\% and compared with cosmological observations. Their observed relative abundance to
hydrogen $H$ is $[1:0.25:3 \times 10^{-5}:2\times 10^{-5}:2\times 10^{-10}]$. The BBN codes calculate these abundances using the laboratory measured nuclear reaction rates, the decay rate of the neutron, the number of light neutrinos and the homogeneous Friedmann-Robertson-Walker (FRW)  expansion of the Universe, as a function of only one variable $\eta\equiv n_B/n_\gamma$, which is the number density fraction of baryons to photons. The present observations are only consistent provided that \cite{burles,olive,particledata} 
 \be\label{eta10}
 \eta_{10}\equiv 10^{10}\eta=6.2\pm 0.6\,, \quad \to \quad \eta = (6.2\pm 0.6) \times 10^{-10}\,.
 \ee
 For mechanisms aimed to explain the origin of the baryon asymmetry see Refs. \cite{BA} and the review \cite{reviewBA}.
Such  a small value indicates that there is about one baryon per $10^9$ photons in the Universe today. Any acceptable theory of baryogenesis should account for such a small number. Furthermore, the present
baryon fraction of the critical density can be calculated from $\eta_{10}$ as
 \be\label{Omega_B}
 \Omega_B h^2 = 3.6271 \times 10^{-3} \eta_{10}=0.0224\pm 0.0024\quad (95\% \,\,\, \mbox{C.L.})
 \ee
This number indicates, as we will discus later, that baryons cannot account for all the matter in the
observed Universe. Concerning the CMB radiation, it certainly represents one of the most remarkable prediction of General Relativity.
The existence of a relic background of photons from Big Bang was predicted by  Gamow and collaborators
(1940), and this fundamental result was obtained accounting for the consistency of primordial nucleosynthesis
with the observed Helium abundance.
The estimation of the temperature background was $T_\gamma\sim 10$ K. A more detailed analysis performed by
Alpher and Herman (1950) \cite{alpher}
led to $T_\gamma \sim 5$ K. It is interesting to point out these results (see \cite{alpher}) slipped into obscurity
since it was not clear if such a radiation would have survived until the present epoch
of the Universe. This problem was again faced about 25 years later by  Dicke, Peebles,
Roll and Wilkinson \cite{wilkinson}, and Penzias and Wilson observed a weak isotropic background signal at a radio
wavelength of 7.35 cm, corresponding to a black-body temperature of $T_\gamma \sim (3.5 \pm 1)$ K.
Since then, many different experiments  confirmed the existence of the CMB.
In particular, the modern experiments are devoted to look for small inhomogenousity of the early Universe which have
distorted the spectrum of the background radiation, leading to the small anisotropies in the temperature.

Despite these crucial predictions and results of Einstein's relativity, the latter appears to be in disagreement with the increasingly high number of observational data (coming for example from SNeIa, large scale structure ranging from galaxies up to galaxy superclusters, CMB) that today are obtained thanks to the achievement of high sensibility of experiments and the advent of the so called {\it Precision Cosmology}. In fact, one of the most exciting discovery of the modern Cosmology is the (strong) evidence of the acceleration of  Universe expansion. This discovery has given rise, from one side, to a more and more growing challenge to understand our Universe,
and, from the other side, it has led to look towards new ideas and theories which go beyond the standard Cosmology and particle physics.

In the attempt to explain the huge amount of recent observational data and more important try to preserve the conceptual structure of General Relativity, cosmologists had the necessity to introduce the two new fundamental concepts of  Dark Matter  and Dark Energy.
A huge  amounts of DM and DE are needed for accounting of the observed cosmic accelerating expansion, formation and stability of large scale structure and, till now, there are no evidence, both experimental and theoretical, able to definitively explain such mysterious components.

The considerable bulk of new observational data  allowed to extend the standard cosmological model to the so called concordance model.
The new picture is essentially the following: the Universe is spatially flat and nowadays  in an accelerating expansion phase.
It is homogenous and isotropic at large scales,  composed by ordinary matter (neutrons, protons, electrons and neutrinos), and filled up  with  DM and  DE. The early phase of the Universe, instead, is characterized by a super-accelerated phase (the so called Inflation) which affected  the anisotropies of CMB and led, via gravitational instabilities, to the formation of large scale structure that  we observe today. After Inflation, the Universe was very hot with high matter density (hot big Bang) which led to the formation of elements via BBN and the CMB radiation.

\subsection{ Cosmic acceleration and the concordance model}

The discovery of the cosmic acceleration has been one of the most important achievements of  modern Cosmology.
It has been realized by studying the luminosity of distant SNeIa. Results showed that the measurements of the SNe luminosity are in agreement with an accelerated expansion of the Universe (in the last 5 Gys), otherwise their luminosity should be about 0.25 mag bigger with respect to the observed one for a decelerating Universe\footnote{It is interesting to recall that initially the observed weak luminosity of SNe was ascribed to a hypothetical (gray) dust present in the Universe, rather to the cosmic acceleration \cite{aguirre,drell}. Only some years later, thanks to the survey of SNeIa (with a red-shift up to $z\simeq 1.8$) performed  with the Hubble Space Telescope, it was definitively confirmed that the weakness of distant SNe is a consequence of the accelerated expansion of the Universe. } \cite{Perlmutter,Riess1998,perl-schmidt,riess2000,caldwell,copeland-sami}.

The determination of the current cosmological scenario follows, besides the Hubble diagram based on distant SN-Ia assumed as standard {\it candle}, also from the analysis of the CMB anisotropies, the power spectrum of large scale structures, and the Hydrogen Lyman-$\alpha$ lines.
In particular, the study of the CMB anisotropies is today one of the more active research lines that cosmologist are pursuing in theoretical framework. The CMB anisotropies in the temperature provided a fingerprint of the Universe before the formation of structures, a period which corresponds to about 380.00 y after the Big Bang. In this period, photons decoupled from baryons and the dynamical evolutions described by the
gravitational field equations predicts the existence of peaks in the angular power spectrum of the CMB temperature fluctuations. BOOMERang \cite{debernardis} and MAXIMA (Millimetric Anisotropy Experiment IMaging Array) \cite{MAXIMA} experiments have been able to detect the position of the first and second peak in the spectrum of the CMB radiation anisotropy. More recently, the NASA WMAP satellite  \cite{bennett,WMAP} has also measured the angular power spectrum of the temperature anisotropies in CMB, concluding definitively that the angular power spectrum is characterized by acoustic peaks which are generated by acoustic waves propagating the primordial fluid of coupled photon-baryon. Moreover, WMAP collaboration together with higher resolution CMB experiment and Galaxy survey 2dFGRS \cite{percival} determined the cosmological parameters with a few percent error.

The positions and the amplitudes of the acoustic peaks are crucial in what they provide many cosmological information about the early Universe.
In fact, they allow to assert that the our Universe is {\it nearly} spatially flat (i.e. the spatial curvature content of the Universe is $\Omega_k\approx 0$, see later). Moreover, combining the CMB measurements with the observational data coming from the large scale structures\footnote{Notice that in such an analysis, one concludes that the parameter characterizing the baryon asymmetry is \cite{dunkley}
 \[
 \eta^{(CMB)}\sim (6.3\pm 0.3)\times 10^{-10} \qquad
 \qquad 0.0215\leq \Omega_B h^2 \leq 0.0239
 \]},
one may conclude that the contribution of the matter to the total matter density is about 25\%. The cluster of Galaxies and the gas fraction in the clusters constraints the matter density density to about 30\%, in which $\sim$ 25\% is due to {\it dark} matter and $\sim $5\% is due to baryonic matter. More precisely,
in terms of the Einstein field equations we have that for a Friedmann-Robertson-Walker Universe one has
 \begin{eqnarray}\label{friedamn}
 H^2 &=& \frac{8\pi G}{3}(\rho_m+\rho_r)+\frac{\Lambda}{3}-\frac{k}{a^2} \\
 &=& H_0^2\left[
 \Omega_r (1+z)^4+\Omega_m (1+z)^3+\Omega_k (1+z)^2 +\Omega_\Lambda\right]\,, \nonumber
 \end{eqnarray}
where $H_0$ is the present value of the Hubble parameter, $\Lambda$ the cosmological constant, $k$ the spatial curvature, $\rho_{m, r}$
the energy density of matter/radiation, $z=\frac{a_0}{a}-1$ the cosmological red-shift, $a$ the scale factor (conventionally it is assumed that today the scale factor assumes the value $a_0=1$), while
the energy density parameters $\Omega_i$, $i=m, r, k, \Lambda$ entering this equation are defined as follows:
 \be\label{Omega}
 \Omega_\Lambda\equiv \frac{\Lambda}{3H_0^2}\,, \quad \Omega_m\equiv \frac{8\pi G \rho_m}{3 H_0^2}\,, \quad
 \Omega_k\equiv \frac{-k}{a^2}\,, \quad \Omega_r\equiv \frac{8\pi G\rho_r}{3H_0^2}\,.
 \ee
The energy density parameters $\Omega_{m, r, k, \Lambda}$ are constrained by the relation
 \be\label{Omega=1}
 \Omega_m+\Omega_r+\Omega_k+\Omega_\Lambda =1\,.
 \ee
According to cosmological models, to get an Universe spatially {\it nearly} flat and in an accelerating phase,
the parameters $\Omega_{m, r, k, \Lambda}$ such that $\Omega_m+\Omega_r+\Omega_k+\Omega_\Lambda =\Omega_0$ must assume
{\it today} the values
 \be\label{Omega-values}
 \Omega_0\simeq 1.02\pm 0.02\,, \quad  \Omega_\Lambda\simeq 0.73\pm 0.04\,, \quad \Omega_m\simeq  0.27\pm 0.04\,, \quad
 \Omega_k\simeq 0\,, \quad \Omega_r\simeq 5 \times 10^{-5}\,.
 \ee
The model we have discussed is the so called {\it concordance model} and is based on above values for $\Omega_{m, r, k, \Lambda}$.

For neutrinos, the corresponding density energy parameter is $\Omega_\nu < 0.015$. Obviously $\Omega_{r, \nu}$ can be neglected. $\Omega_m$ in (\ref{Omega-values}) contains the baryonic contribution ($\Omega_b$) to the total matter content in the Universe. Data gives
 \[
\Omega_b\simeq 0.044\pm 0.004\,,
  \]
a values consistent with the ratio of baryons to photons $\eta\simeq (6.1\pm 0.7) \times 10^{-10}$. $\Omega_b$ does contain all forms of radiant baryonic mass, which are stair-light ($\Omega_{star}\simeq (1-2) \times 10^{-3}$), gas and stars remnant in Galaxies ($\Omega_{\text{gas-stars remnant}} < 10^{-2}$). Notice that the intergalactic gas contains:
 1) hydrogen clouds and filaments (observed via Ly$\alpha$ absorption);

 2) warm gas in group of Galaxies (radiating soft $X$-rays;

 3) hot gas in clusters (observed in KeV $X$-rays).

This analysis yields the conclusion that there is a missing fraction in matter content which does not radiate but manifests its presence only via
gravitational interaction. The missing matter is therefore called DM:
 \[
 \Omega_{DM} = \Omega_m - \Omega_b \simeq 0.23 \pm 0.05\,.
 \]
It is remarkable that the missing density matter is larger with respect to the know form of baryonic matter.

As before mentioned, BOOMErang, MAXIMA and WMAP experiments have determined that the Universe is close and spatially flat, in agreement with the concordance model.The latter has been further confirmed by the detection detection of the polarization fluctuations by DASI \cite{kovac} and of the temperature polarization angular power spectrum by WMAP \cite{kogut}. Recent works have found evidences of positive cross-correlations between the CMB temperature map and the large scale structure distribution of Galaxies \cite{boughn}. These work are very relevant in what they provide an evidence of the existence and {\it dominance}  of dark energy in the recent history of the Universe evolution.

Although the concordance model is the best model we have that fit CMB data as well as other important cosmological data (BBN, survey of large scale structure of Galaxies, SN Ia magnitude red-shift diagram, measurements of the Hubble constant), there are still many open issues related to observations and their interpretations.
In this review we discuss new ideas related to DM and DE, and their interpretation in the perspective of the modern particle physics and cosmology.

\section{Dark Energy}

According to General Relativity, a Universe filled with ordinary matter or radiation, is decelerated as the mutual gravitational attraction opposes to expansion. Therefore, to explain the observed accelerating phase of the Universe, cosmologists need to introduce the concept of Dark Energy, a unknown form of energy responsible of the increasing rate of expansion of the Universe.

The simplest forms for DE is the cosmological constant $\Lambda$, which is, in a nutshell, a constant energy density filling space homogeneously, and physically it is equivalent to vacuum energy. On the other hand, one may also have that the {\it cosmological constant} actually is not a constant, but have a dynamics. In such a case, one introduces scalar fields such as quintessence or moduli, dynamic quantities whose energy density can vary in time and space.

The Einstein field equations, and more generally, many cosmological models impose that the evolution of the expansion rate is parameterized by the cosmological equation of state. The general expression is
 \[
 p=w \rho\,,
 \]
where $w$ is the adiabatic index. Measuring the equation of state of DE is one of the biggest efforts in observational cosmology today. In the next subsections, we shall review some models

Adding the cosmological constant to the  FRW standard cosmology one obtains the so-called $\Lambda$-CDM model. This is the best cosmological model owing its precise agreement with observations.

The fact that both cosmic speed-up and DM is understood as a signal of a possible breakdown in our understanding of gravitation laws open the possibility that gravity, as described by General Relativity, could be not a complete theory, and modifications/generalizations are therefore required. On the other hands, although General Relativity has been very well tested at scales of solar system, its validity has never been tested at scales larger than solar system, i.e. cosmological and astrophysical scales, and therefore it is not excluded a priori that it could be broken at these scales \cite{will}.
Although the modifications or generalizations of the gravitational sector, i.e. of General Relativity, are difficult and not so immediate, there exist in literature numerous models. However, most of them turn out to be non-viable \cite{will}, and the most well known alternative approach to General Relativity are, essentially, the following: the scalar-tensor theory
\cite{maeda,bergmann,brans-dicke,faraoni2004a,nordtvedt,wagoner}, the Dvali- Gabadadze-Porrati gravity \cite{dvali}, the braneworld gravity \cite{maartens2004}, the Tensor-Vector- Scalar \cite{bekenstein2004}, the Einstein-Aether theory \cite{Jacobson-Mattingly}.

For recent reviews on models of DE, see \cite{silvestri,turner,durrer,caldwell,sami,copeland-sami,capozzielloPR}.

\subsection{The cosmological constant $\Lambda$}

According to particle physics, the vacuum energy turns out to be the result of the vacuum fluctuations of the quantum fields associated to particles, see for example \cite{zeldovich,weinberg-cosm-constant,padmanabhan,bousso,carroll}.
 The vacuum density energy of a free quantum field is defined as
 \[
 \rho\sim \int d^3 {\bf k} \, \omega_k \sim \int_0^\infty dk\, k^2\, \sqrt{k^2+m^2}\,, \quad \omega_k=\sqrt{k^2 +m^2}\,.
 \]
This quantity is evidently divergent in the regime of arbitrarily very small wavelength (or equivalently large wave number $k$). These divergences reflect the fact that we are assuming that our description of the physical words does work at these scales, where gravity might play a very relevant role: However, in absence of a complete theory of quantum gravity, this assumption is not reasonable, and therefore an estimation of the vacuum density energy can be obtained
by introducing a cutoff on momenta/energy: $k_{Pl}=E_{Pl}=m_{Pl}=(8\pi G)^{-1/2}\sim 10^{18}$GeV (in natural units). As a result, one obtains $\rho^{\text{(th)}}\sim m_{Pl}^4\sim 10^{72}$GeV$^4$. Comparison this value with that one inferred in cosmology to explain the cosmic acceleration, $\rho^{(\text{exp})}\lesssim 10^{-48}$GeV$^4$, one gets  $\rho^{\text{(exp)}}\sim 10^{-120} \rho^{\text{(th)}}$ (this is the well known cosmological constant problem).

From a theoretical point of view, the interesting aspect related to the cosmological constant is that the pressure and the energy density are related as\footnote{The fact that $\Lambda$ may imply a negative pressure $p$ is related to the following classical thermodynamics arguments: to do work on a container (Universe), the energy must be lost from inside a container. Denoting with $V$ and $p$ the volume and the pressure, respectively, then a change in volume $dV$ requires the work $- p dV$. The amount of energy in a box of vacuum energy actually increases when the volume increases ($dV > 0$), because the energy is equal to $\rho V$, where $\rho$ is the energy density of the cosmological constant. Therefore, $p$ is negative and, in fact, $p = -\rho$ .} $p=-\rho$, i.e. one obtains a  {\it negative} pressure which represents a necessary condition to cause the accelerated expansion of the Universe. This can be immediately seen by writing the energy-momentum tensor associated to the cosmological constant.
 \[
 T^{(\Lambda)}_{\mu\nu}= - \Lambda\, g_{\mu\nu}\,.
 \]

Related to the problem of the cosmological constant there are some further considerations that deserve to be mentioned.

In the framework of Supersymmetry (SUSY),
the problem of zero point energy is naturally solved. In fact, since boson and fermion degrees of freedom contribute with opposite signs
to the vacuum energy, the total vacuum energy vanishes (supersymmetric theories do not admit a non-zero cosmological
constant!). However, we do not live in a supersymmetric Universe as arises from experiments and observations. Therefore, if SUSY is a symmetry that realized in nature, it must be broken at some scale which is greater than $M_{\text{SUSY}}\sim $TeV (this scale is relevant to solve the hierarchy problem in SUSY models). However, also using $M_{\text{SUSY}}\sim $TeV, we are still far away from the observed value of the energy density of the Universe by many orders of magnitudes. We do not know how Planck scale or SUSY breaking scales are related to the
observed vacuum scale. Experiments at LHC performed at CERN are searching signals able to confirm the existence of SUSY particles, a result that might shed light in a near future on the vacuum energy problem.

Another feature related to the cosmological constant problem (and hence the cosmic acceleration) is the following: why at the present epoch the energy density of the Universe in the form of cosmological constant and the matter energy density have the same magnitude? This is  {\it the coincidence problem}. At the moment, no satisfactory answers or models have been proposed to solve this problem.

Finally, we mention the interconnection between and string theory. The latter predicts a landscape of many (about $10^{500}$ de-Sitter vacua - see \cite{bousso} and references therein). The anthropic principle allows to select one of this vacua, which {\it contains} our Universe. For more details and explanations, see Refs. \cite{copeland-sami,anthropic}.

\subsection{Quintessence}

Quintessence is a scalar field responsible of the accelerated phase of the Universe, and represents the simplest field that can provide the missing energy. If the vacuum energy does not vary with space or time, hence it is not dynamical, then one recovers the cosmological constant before discussed. To obtain an {\it effectively} dynamical vacuum energy, one has to
introduce a new degree of freedom, a scalar field $\phi$ \cite{watterich,ratra-peebles,frieman,zlatev,silvestri,turner,durrer,caldwell,copeland-sami}, whose Lagrangian density is
 \[
 {\cal L}=\frac{1}{2}\partial_\mu \phi \partial^\mu \phi - V(\phi)
 \]
This Lagrangian density is a particular case of a more general written in the framework of scalar-tensor theories \cite{bergmann,brans-dicke,faraoni2004a,nordtvedt,wagoner,maeda}, whose action is of the form
 \begin{equation}\label{scalar-tensor-action}
  S_{scalar-tensor}=\frac{1}{8\pi G}\int d^4 x \sqrt{-g} \left[
  \Phi(\phi) R -\frac{h(\phi)}{2}\partial_\mu \phi \partial^\mu \phi -v(\phi)\right]+S_{matter}\,.
 \end{equation}
The form of the energy-momentum tensor implies that (assuming $\phi$ homogeneous, $\phi=\phi(t)$)
 \[
 \rho=\frac{\dot \phi}{2}+V\,, \qquad p=\frac{\dot \phi}{2}-V
 \]
The evolution of the scalar field follows from the Klein-Gordon equation of motion ${\ddot\phi}+
+ 3H{\dot\phi} + dV/d\phi = 0$. The equation-of-state parameter is therefore
 \[
 w = \frac{-1+{\dot \phi}^2/2V}{+1+{\dot \phi}^2/2V}
 \]
This quantity describes a scalar-field as DE, in the sense that
in the regime ${\dot \phi}^2/2V \ll 1$, one has $w=-1$ i.e. the scalar field evolves slowly
and it behaves like a slowly varying vacuum energy characterized by $\rho_{vacuum}\sim V$. In general, $w$ varies with time ranging from
the value $w=-1$ (rolling very slowly) to $w=+1$ (evolving very rapidly)\footnote{Many scalar-field models can be classified dynamically as {\it thawing} or {\it freezing} \cite{caldewell-linder}. In the first case, the field rolls more slowly as time progresses, i.e., $dV/d\phi\ll 3H{\dot \phi}$ in the equation of motion for $\phi$ (this can happen if, for instance, $V$ falls off exponentially or as an inverse power law at large $\phi$), whereas in the second case,  at early times the field is frozen by the friction term and acts as vacuum energy -
when the expansion rate drops below $H = \sqrt{d^2V/d\phi^2}$, the field begins to roll and $w$ evolves away from -1 (an example of a thawing model is a massive scalar field, with $V =  m^2
\phi^2/2$) \cite{caldewell-linder,turner}.}.

An interesting extension of models involving scalar fields as DE is represented by
model in which non-canonical kinetic terms are present in the Lagrangian. Such a model are called k-essence. The Lagrangian density is of the form
 \[
 {\cal L}(\phi, X)= K(X)-V(\phi)\,, \qquad X\equiv \frac{1}{2}\partial_\mu \phi\partial^\mu \phi\,,
 \]
where $K$ is a positive semidefinite function. The stress tensor yields
 \[
 \rho=2 X \frac{dK}{dX}-K+V\,, \qquad p=K-V\,,
 \]
from which it follows
 \[
 w=\frac{K-V}{2X dK/dX-K+V}\,.
 \]
Consequences of these models have been largely studied in literature \cite{k-essence}.

\subsection{Mass varying neutrinos}

The possibility that DE couple with fermion sector has been explored quite recently in \cite{fardon,peccei-massvaryingnu}. The idea relies on the fact that the neutrino mass (more generally the neutrinos masses of each flavors) is not a constant, but depends on the neutrino number density $n_\nu$ (and therefore with the temperature \cite{peccei-massvaryingnu}). A mass-varying neutrinos immediately can be related to a slowly-varying scalar field (like quintessence), denoted in literature {\it acceleron}, whose value determines the neutrino mass $m_\nu$. The main lines of this idea are the following. In a non-relativistic regime, the energy the energy density of a fluid of neutrino DE is given by
 \[
 \rho_{DE}=m_\nu n_\nu + \rho_a (m_\nu)\,,
  \]
where $\rho_a (m_\nu)$ is the acceleron density. By using the conservation of the energy $\nabla_\mu T^{0\mu}_{DE}=0$, $T^{\mu\nu}=diag(\rho_{DE}, p_{DE}, p_{DE}, p_{DE})$ and the condition
 \[
 \frac{\partial \rho_{DE}}{\partial m_\nu} = n_\nu+\frac{\partial \rho_a}{\partial m_\nu}\,,
 \]
one finds
 \[
 w=-1 + \frac{m_\nu n_\nu}{m_\nu n_\nu+\rho_a}\,.
 \]
leading to the value $w \simeq -1$. Consequences of this model in the framework of neutrino oscillations have been investigated in \cite{kaplan-nelson-weiner}.

Although this model is very interesting, it is not immune from severe difficulties related to
instability to the growth of perturbations that renders it unsuitable for explaining cosmic acceleration (the gradient-energy density turns out to be too small to prevent the growth of spatial fluctuations, implying that $c^2_s = w < 0$, which gives rise to a dynamical instability to the rapid growth of perturbations to the mass varying neutrinos energy density - such instabilities are similar to ones coming in the attempt to couple dark matter and dark energy \cite{instabilitymvaryingmassnu,anderson-carroll}) \cite{brookfield}.

In any viable model in which scalar fields play the role of DE, it is required
that the mass of the field is $\sim 10^{-42}$ GeV, and the field must have an amplitude of $10^{19}$ GeV. Moreover, these models should be also able to explain the coincidence problem.
Despite these non trivial difficulties, many models have been proposed. Here we just mention some of them: the phantom model (as follows from observations, there is the possibility that the adiabatic parameter $w$ may assume values $< -1$, meaning that DE violates the null dominant energy condition) \cite{phantom}, coupled DE model (DE and DM interact with each other or with ordinary matter - in these models, the coupling between dark matter and dark energy generally gives rise to instability \cite{instabilitymvaryingmassnu,anderson-carroll}.) \cite{coupled,instabilitymvaryingmassnu,anderson-carroll}, cosmic axion (or pseudo Nambu Goldstone boson with extremely low mass) \cite{cosmicaxion}, tracker fields (in SUSY version of QCD, point-like scalar fields are generated by condensation of hidden sector quark-antiquark pairs) \cite{trackerfields}, spintessence (a complex field $\phi= R e^{i\Theta}$ spinning in a U(1)-symmetric potential $V(|\phi|)$) \cite{spintessence}, ghost condensate \cite{ghostcondensate}, etc.

\subsection{$f(R)$ theories of gravity}

In the last years, among the different approaches proposed to generalize Einstein's General Relativity, the so called $f(R)$-theories of gravity received a growing attention. The reason relies on the fact that they allow to explain, via gravitational dynamics, the observed accelerating phase of the Universe, without invoking exotic matter as sources of dark matter or extradimensions.
In these models, the {\it breakdown} of the Einstein theory consists in a geometrical generalization of the gravity action, i.e. the gravity Lagrangian for these theories is a generic function of the Ricci scalar curvature $R$. Therefore, the well known Hilbert-Einstein Lagrangian, linear in the Ricci scalar R (see, for example, the recent review \cite{faraoni})
 \[
 S_{HE}=\frac{1}{8\pi G}\int d^4 x \sqrt{-g} \, R + S_{matter},,
 \]
is generalized as
 \begin{equation}\label{Jordan-frame}
 S=\frac{1}{8\pi G}\int d^4 x \sqrt{-g} \, f(R)+S_{matter}\,,
 \end{equation}
It is worth noting that the presence of higher order terms into gravity Lagrangian are predicted in many contexts, such as, for example, one-loop corrections in the procedures of field quantization on curved spacetimes \cite{birrell}, in any perturbative approach aimed to achieve a self-consistent theory of quantum gravity, in the low energy limit of string/M-theory \cite{odintsov-nojiri-plb}.

An interesting feature of this model is that $f(R)$ theories of gravity are equivalent to scalar-tensor theories \cite{chiba}. To show this statement, consider the standard action of gravity with a scalar field (here denoted $\zeta$)
 \begin{equation}\label{F-zeta}
 S_\zeta = \frac{1}{8\pi G} \int d^4 x \sqrt{-g} \left[F(\zeta)-\frac{dF}{d\zeta}\, (R-\zeta)\right]
 +S_{matter}
 \end{equation}
The equation of motion for the scalar field $\zeta$
 \[
 \partial_\mu \left(\frac{\partial {\cal L}}{\partial (\partial_\mu\zeta)}\right)=
   \frac{\partial {\cal L}}{\partial \zeta}\,,
 \]
gives $R-\zeta=0$ provided $d^2F/d\zeta^2=0$. This result shows the equivalence of $S_\zeta$ with $S\sim \int d^4 x \sqrt{-g} \, f(R)+S_{matter}$ (with $f\equiv F$), and therefore the action (\ref{F-zeta}) and (\ref{scalar-tensor-action}) are equivalent if
 \[
 \Phi(\zeta)=\frac{dF}{d\zeta}\,, \qquad V(\zeta)=-F+\zeta \frac{dF}{d\zeta}\,, \qquad
 h=0\,.
 \]
The latter condition implies that the kinetic term vanishes in the density Lagrangian.
It is interesting to observe that with an appropriate conformal transformation, one can recast the above action in the form of Hilbert-Einstein with a scalar field. In fact,
to pass from the Jordan frame (the action (\ref{Jordan-frame})) to the Einstein-frame (see Eq. (\ref{Einstein-frame}) below), one uses the conformal transformation
 \[
  g_{\mu\nu}\to g_{\mu\nu}^E=\frac{d\Phi}{d\phi}g_{\mu\nu}\,,
  \qquad
 \frac{dF}{d\zeta}=\exp{(\sqrt{16\pi G/3}\, \phi)}\,.
 \]
It follows
 \begin{equation}\label{Einstein-frame}
  S = \int d^4 x \sqrt{g^E} \left[\frac{1}{16\pi G}\, R^E-\frac{1}{2}g_{\mu\nu}^E \partial^\mu \phi
  \partial^\nu \phi-V(\phi)\right]\,,
  \end{equation}
where
 \[
 V(\phi)\equiv \frac{\zeta(\phi)F'(\zeta(\phi))-F(\zeta(\phi))}{16\pi G [F'(\zeta(\phi))]^2}\,,
 \qquad F'=\frac{dF}{d\zeta}\,.
 \]
A comment is in order. By analyzing scalar-tensor theories of gravity  in the Einstein frame, it arises that they look like General Relativity with the {\it canonical} scalar field $\phi$. The substantial difference relies on the fact that the metric $g^E_{\mu\nu}$ is {\it not} the metric whose geodesics determine particle orbits. The latter is determined by the Jordan-frame metric $g_{\mu\nu}$. Thus, scalar-tensor theories in the Einstein frame resemble General Relativity with an extra, non-geodesic, force on the particle \cite{caldwell}. These may also be generalized by chameleon theories \cite{khoury-weltman}, where the scalar coupling to matter may differ for different matter fields. Viewed in the Jordan frame, scalar-tensor theories are characterized by the fact that one has the usual two propagating tensorial degrees of freedom and in addition there is a new propagating scalar degree of freedom.

The field equations for a $f(R)$-theory are (see for example \cite{capozziello-jcap})
 \[
 f' R_{\alpha\beta}-\frac{1}{2}\, g_{\alpha\beta} f = (g_{\alpha \mu}g_{\beta\nu}-g_{\alpha \beta}g_{\mu\nu}) \nabla^\mu\nabla^\nu f'+ T_{\alpha\beta}^{{\text matter}} \,, \quad f' = \frac{df}{dR}\,.
 \]
This equation can be recast in the Einstein-like field equation
 \[
 R_{\alpha\beta}-\frac{1}{2}\, g_{\alpha\beta} R=\frac{1}{f'} T_{\alpha\beta}^{{\text matter}}
  + T_{\alpha\beta}^{{\text curv}}\,,
 \]
in which we see that matter couples non-minimally to geometry through the term $1/f'$ and we have defined {\it curvature} stress–energy tensor as
 \[
 T_{\alpha\beta}^{{\text curv}} = \frac{1}{f'}\left[\frac{1}{2}\, g_{\alpha\beta} (f-Rf')+
 (g_{\alpha \mu}g_{\beta\nu}-g_{\alpha \beta}g_{\mu\nu}) \nabla^\mu\nabla^\nu f'\right]\,.
 \]
Notice that $\nabla^\alpha T_{\alpha\beta}^{{\text curv}}=0$.

From these equation it follows that the cosmological dynamics is determined by its
energy budget determined by the ordinary matter and the curvature contribution. In particular, for a FRW metric, the cosmic acceleration is achieved when the right handed side of the acceleration equation remains positive
 \[
 \frac{\ddot a}{a}\propto -(\rho_{tot}+p_{ tot})
 \]
where
 \[
 \rho_{tot}=\rho_{ matter}+\rho_{ curv}\,, \quad
  p_{ tot}=\rho_{ matter}+p_{ curv}\,.
  \]
In particular, for a dust ($p_{ matter}=0$) dominated model one has
 \[
 \rho_{matter}+\rho_{ curv}+3p_{curv}<0 \quad
 \to \quad w_{curv} < -\frac{\rho_{ matter}+\rho_{ curv}}{3\rho_{ curv}}\,,
  \]
where
 \[
 \rho_{curv}=\frac{1}{f'}\left[\frac{f-Rf'}{2}-3H {\dot R} f''\right]\,,
 \]
 \[
 p_{curv}={\dot R}^2 f'''+2 H {\dot R} f'' + {\ddot R} f'' +\frac{1}{2}(f-Rf')\,,
 \]
and
 \begin{eqnarray}
 w_{curv}&=&\frac{p_{curv}}{\rho_{curv}} \nonumber \\
 & = &-1+\frac{{\ddot R}f''+{\dot R}({\dot R}f'''-Hf'')}{\frac{1}{2}(f-Rf')-3H{\dot R}f''}\,.
 \nonumber
 \end{eqnarray}
Of course, due to the freedom in choosing the explicit form of the function $f(R)$, many models have been investigated in literature\footnote{In models studied in literature, it is assumed that $f' > 0$ in order that the effective gravitational coupling is positive and $ f'' > 0$ to avoid the Dolgov-Kawasaki instability \cite{dolgov-kawasaki,faraoni2006}}. We shall analyze some of them.

 \begin{itemize}
   \item The models proposed in \cite{capozziello1,carroll1} achieved late-time acceleration of the Universe
by choosing
 \begin{equation}\label{f(R)n}
 f(R)= R + a R^{-n}\,,
  \end{equation}
with $a=constant$ and $n \geqslant 1$. However, such a kind of models, as was realized in \cite{f(R)instable,dolgov-kawasaki,hu}, possess instabilities which, among other problems, prevent them from
having a matter dominated epoch (see also \cite{altri-f(R)}).

   \item As discussed in \cite{hu,silvestri}, the choice of the function $f(R)$ must satisfies
the phenomenology dictated by the $\Lambda$CDM model. This is an important condition. Following this guide-line, and assuming that $f(R)=R+g(R)$, the criteria that $g(R)$ must obey are \cite{hu,silvestri}
 \[
 \lim_{R\to \infty}g(R)=\Lambda\,,\quad
 \lim_{R\to 0}g(R)=0\,.
 \]
The choice of the generic function $g(R)$ is, as proposed in \cite{hu,silvestri}
 \[
 g(R)=-m^2 \frac{c_1(R/m)^n}{c_2(R/m)^n+1}\,,
 \]
where $c_{1,2}$ are dimensionless parameters and $m^2=\kappa^2
\rho_0/3$, being $\rho_0$ the average density today. In the high
curvature regime, such a model implies an action proportional to
$\int d^4x \sqrt{-g} (R+\Lambda)$.
Stringent the constraints on the these free parameters come from solar and
galaxy systems on the first derivative of $g(R)$. They are
\begin{equation}\label{bound}
\Delta g_R \equiv |g_R (R)-g_{R\infty}| < 10^{-\delta}
\end{equation}
where $\delta=11$ in a model in which the sun is embedded in a
medium at cosmological density, whereas extrapolations of galaxies
rotation curve measurements give $\delta=6$. $R_\infty$ is the
value of $R$ at large radial distance from the bound system, and
$g_R=\partial g/\partial R$.
   \item Recent investigations \cite{tsujikawa} have shown that for
a $f(R)$ dark energy model with Lagrangian density $f(R)\sim
\alpha R^n -\Lambda$, with $n=1+m$ and $m=Rf_{, RR}/f_{, R}$, the
severest constraint on $m$, hence on $n$, comes from laboratory
experiments. The latter in fact, tell that a strong deviation from
General Relativity has not been observed up scales $\sim 1 $mm.
This means that $m\vert_{today} \lesssim 10^{-58}$. As a
consequence, the constraint on $n$ becomes $n\lesssim 1+10^{-58}$.
  \item Stringent constraints on parameters entering the form of $f(R)$ come from the
PPN approximation (i.e. such models must not violate the experimental constraints on Eddington parameters \cite{will}) - see \cite{capozziello-troisi,faraoni-2006,faraoni} and references therein.

 \end{itemize}

In conclusion of this paragraph, we wish to point out that higher order gravity theories
are from a side able to account for the cosmic acceleration of the Universe expansion, both in the late and in the early Universe \cite{starobinsky,kerner}, and from the other side, they play also a relevant role at astrophysical scales. In fact, the modification of the gravity Lagrangian can affect the gravitational potential in the low energy limit \cite{stelle} which
reduces to the Newtonian potential on the solar system scale. Such a modified
gravitational potential offer the possibility of fitting galaxy rotation curves without
the need of dark matter. We shall not discuss these interesting results. The interested reader may refer to \cite{capozziello-rotationcurve,milgrom,f(R)-DM}.

For a complete review on $f(R)$ theories of gravity, and their application in cosmology and astrophysics, see in particular \cite{faraoni,silvestri}, and references therein.

\subsection{Non-Linear Electrodynamics}

With the aim to build up a classically singularity-free theory of the electron, that is a theory in which infinite physical quantities are avoided, Born and Infeld \cite{bohr} proposed a model in which additional terms or modifications of the standard electrodynamics were included. To prevent the infinite self energy of point particles (as follows from standard electrodynamics), they introduced an upper limit on the electric field strength and considered the electron as an electric particle with finite radius. In successive investigations, other examples of nonlinear electrodynamics Lagrangians were proposed by Plebanski, who also showed that Born-Infeld model satisfy physically acceptable requirements \cite{plebanski}.

Nonlinear electrodynamics represents an interesting class of models ables to account for accelerated expansion of the Universe. Following the modification of the dynamics of
the gravitational field in the low-curvature regime (such a
modification can be achieved by considering in five-dimensional
scenarios the effects of the bulk on the dynamics
of gravitation on the brane \cite{deffayeta}, or by directly adding to
the four-dimensional gravitational action terms with negative
powers of the curvature scalar - Eq. (\ref{f(R)n}) with $n=1$) - it was recently proposed a model in which the action for the electromagnetic field is that of Maxwell with an extra term \cite{novello}
  \begin{equation}\label{S-NLE}
  S=\int d^4x \sqrt{-g} \left[-\frac{X}{4}+\frac{\gamma}{X}\right]
  =\int d^4x \sqrt{-g} L^{NLE}(X) \,, \qquad X\equiv F_{\mu\nu} F^{\mu\nu}\,.
  \end{equation}
This action is gauge invariant, so that for construction the charge conservation is guaranteed. The model involves only the electromagnetic field, without invoking scalar fields or more speculative ideas related to higher-dimensions and brane worlds. The action (\ref{S-NLE}) is only an example of
a more general class of actions for the electromagnetic field with Lagrangians that can be written as
 \[
 L=\sum_k c_k X^k\,,
 \]
where the sum may involve both positive and negative powers of $X$.

Due to the isotropy of the spatial sections of the FRW model, an average procedure
is needed if electromagnetic fields are to act as a source of gravity. Let us define the volumetric spatial average of a quantity $X$ at the time $t$ by \cite{novello}
 \[
 {\bar X}=\lim_{V\to V_0}\frac{1}{V} \int d^3 x \sqrt{-g} X \,, \qquad V=\int d^3 x \sqrt{-g}\,.
 \]
$V_0$ is a sufficiently large time-dependent three-volume. In this notation, the electromagnetic
field can act as a source for the FRW model if
 \[
 {\bar E}_i = {\bar B}_i =0\,, \qquad {\bar E}_i{\bar B}_j =0\,,
 \]
 \[
 {\bar E}_i{\bar E}_j=\frac{E^2}{3}\, g_{ij}\,, \qquad {\bar B}_i{\bar B}_j=\frac{B^2}{3}\, g_{ij}\,.
 \]
Moreover, it can be shown that \cite{novello}
 \[
 \rho=-L-4E^2 \frac{\partial L^{NLE}}{\partial X}\,, \qquad\qquad
 p=L+\frac{4}{3}(E^2-2B^2)\frac{\partial L^{NLE}}{\partial X}\,.
 \]
The relevant case studied in cosmology is that one for which $E=0$. Since we are assuming that ${\bar B}_i = 0$, the magnetic field induces no directional effects in the sky, in
accordance with the symmetries of the standard cosmological model.

The equation of states following from the action (\ref{S-NLE}) are \cite{novello}
 \[
 p=\frac{1}{3}\,\rho\,, \qquad \qquad p=-\frac{7}{3}\,\rho\,.
 \]
It is precisely the latter equation of state with
negative pressure that may drive the acceleration of the Universe (for further details, see \cite{novello}).

Consequences of nonlinear electrodynamics have been studied in many contexts, such a, for example, cosmological models \cite{cosmology}, black holes and wormhole physics \cite{BH,wormhole}, and astrophysics
\cite{applications,NLE-atri}.

\subsection{Dvali-Gabadadze-Porrati (DGP) Gravity}

The action of DGP gravity postulates a (4+1)-dimensional Universe in which the bulk of the five-dimensional spacetime is the Minkowski space with an embedded (3+1)-dimensional brane (our Universe) on which matter fields live \cite{dvali}. The action reads
  \[
  S=\int d^5 x \sqrt{-g}\frac{R}{16\pi G^{(5)}}+\int d^4 x \sqrt{-g^{(4)}}\left[
  \frac{R^{(4)}}{116\pi G}+L_{matter}\right]
  \]
where $G^{(5)}$ is the $5-D$ gravitational constant, $g$ and $R$ are the $5-d$ metric determinant and Ricci scalar, $g^{(4)}$ and $R^{(4)}$ the induced metric determinant
and Ricci scalar on the brane. The DGP Friedmann equation are \cite{deffayet,caldwell}
 \[
 H^2\pm \frac{H}{r_0}=\frac{8\pi G}{3}\, \rho\,.
 \]
where $r_0=G^{(5)}/2G$. The minus sign in this equation implies that at early times, when $H \gg r_0$ the usual Friedmann equation is recovered, instead when $H$ decreases, then the new term kicks in, so that $H \to r_0^{-1}$ at late times, i.e., the
Universe asymptotes at late times to a de Sitter phase.

Many other models have been proposed in literature. Here we just recall some of them not discussed in this paper (for an exaustive and complete list see the reviews \cite{caldwell,copeland-sami,silvestri}):
\begin{itemize}
  \item DE models that involve a fluid known as a Chaplygin gas (this fluid also leads to the acceleration of the Universe at late times, and in its simplest form has the following specific equation of state:
 $p =A/\rho$ where A is a positive constant)\cite{moschella,copeland-sami,altri-chaplygin-gas}.
  \item DE has been used as a crucial ingredient in a recent attempt to formulate a cyclic model for the Universe \cite{baum}.
  \item In Ref. \cite{capozziello-mercadante} it has been proposed a model based on the concept of {\it Dark Metric}, which could completely bypass the introduction of {\it disturbing} concepts as Dark Energy and Dark Matter.
  \item Degravitation of the vacuum energy - The idea of the degravitation relies on the replacement of the Einstein field equations by
      \[
      G^{-1}(L^2 \Box ) G_{\mu\nu}= 8\pi T_{\mu\nu}\,,
      \]
where $L$ is a  characteristic scale such that $G\to G^{(0)}$ for $L^2 \Box\to \infty$, and $G\to 0$ for $L^2\Box\to 0$. This model offers a dynamical solution to the cosmological
constant problem (it indeed allows for a large cosmological constant) \cite{degravitation}.
 \item The physics beyond the standard model and cosmology implied by cosmic acceleration may have other observable/experimental consequences, apart simply from its effect on cosmic expansion. In this context, searching for a possible violation of the Lorentz invariance violation will be extremely relevant \cite{caldwell,lorentzviolatin}.
  \end{itemize}

\subsection{Massive Gravity}

Recently, de Rham, Gabadadze and Tolley (dRGT) \cite{derham,derhamPRL} have proposed a model of gravity in which the mass of the graviton
is taken into account. First attempts in this direction were proposed long time ago by Fierz and Pauli \cite{pauli},
who constructed a (ghost-free) linear theory of massive gravity. As later realized, the Fierz-Pauli theory is affected by some pathologies
as, for example, it is in conflict with solar system tests \cite{solartest,vandam}. A renew interest for this theory is arisen in the last years
thanks to the Stueckelberg formalism introduced in \cite{siegel,arkani,kurt}.
The dRGT model relies on the idea to add higher order self-interaction graviton terms to the Einstein-Hilbert action in order to
get rid of the Boudware-Deser instability \cite{deser}.

The action of the massive gravity is described in terms of the usual metric $g_{\mu\nu}$ and the four scalar field $\phi^a$, $a = 0, 1, 2, 3$ called the St\"uckelberg fields \cite{derham}
 \begin{equation}\label{action}
    S=M_P^2 \int d^4 x \sqrt{-g} \left[\frac{R}{2}-\frac{m_g^2}{8}{\cal U}(g_{\mu\nu}, {\cal K}_{\mu\nu})\right]+S_m\,,
 \end{equation}
where $M_P^2 = (8\pi G)^{-1} \simeq 10^{19}$GeV is the Planck mass (in natural units), $m_g$ the graviton mass, and ${\cal K}_\mu^\nu=\delta_\mu^\nu-(\sqrt{\Sigma})_\mu^\nu$, with
$\Sigma_{\mu\nu}\equiv \partial_\mu \phi^a \partial_\nu \phi^b \eta_{ab}$. $S_m$ is the action for matter field.
The explicit form of the mass term in (\ref{action}) is given by
 \begin{equation}\label{calU}
    {\cal U}={\cal U}_2+\alpha_3 {\cal U}+\alpha_4 {\cal U}\,,
 \end{equation}
where $\alpha_3$, $\alpha_4$ are constants, and
 \[
 {\cal U}_2 = [{\cal K}]^2-[{\cal K}^2]\,, \quad {\cal U}_3 = [{\cal K}]^3-3[{\cal K}][{\cal K}^2]+2[{\cal K}^3]\,,
 \quad {\cal U}_4 = [{\cal K}^4]-6 [{\cal K}]^2[{\cal K}^2]+8[{\cal K}^3][{\cal K}]-6[{\cal K}^4]\,.
 \]
Here the symbol $[...]$ stands for the trace of the matrix, i.e. $[{\cal K}]$= Tr ${\cal K}$, and so on.
The Einstein field equations following from the action (\ref{action}) read
\begin{equation}\label{eisteineq}
    G_{\mu\nu}=\frac{8\pi}{M_P^2}\, (T_{\mu\nu}^{(m)}+T_{\mu\nu}^{(\cal K)})\,,
\end{equation}
where $G_{\mu\nu}$ is the usual Einstein tensor, $T_{\mu\nu}^{(m)}$ the usual energy momentum tensor of matter fields, and
$T_{\mu\nu}^{(\cal K)}$ the effective energy momentum tensor associated to the potential $m^2_g {\cal U}$.
In what follows we shall assume that the early primordial plasma is described by a perfect fluid so that
$T_{\quad 0}^{(m)\,0}=\rho_m$, $T_{\quad i}^{(m)\, j}=-p_m \delta_i^j$, and the continuity equation holds
$\nabla_\mu T^{(m)\, \mu\nu}=0$, or ${\dot \rho}_m + 3H (\rho_m +p_m)=0$ in a (flat) Friedman-Robertson-Walker Universe $ds^2 = dt^2-a^2(t) (dx^2+dy^2+dz^2)$.
We do not report here the explicit expression of the tensor $T_{\mu\nu}^{(\cal K)}$ (see for example \cite{gong,gratia,langlois}). What is relevant for our aim is that $T_{\mu\nu}^{(\cal K)}$ can be written in the form of a perfect fluid, $T_{\quad 0}^{({\cal K}) \,0}=\rho_g$, $T_{\quad i}^{({\cal K}) \, j}=-p_g \delta_i^j$, with
${\dot \rho}_{\cal K} + 3 H (\rho_{\cal K} +p_{\cal K})=0$, where $\rho_g$ and $p_g$ represent an effective energy and pressure density arising from massive gravity.
Depending on some specific choice of some functions of the theory (branches of solutions), both $\rho_g$ and $p_g$ assume a particular form.
A solution is  \cite{langlois}
 \begin{eqnarray}\label{rhog2}
 \rho_g &=& -m^2_g M_P^2 \left(1-\frac{H}{H_c}\right)\left[6+4\alpha_3+\alpha_4-(3+5\alpha_3+2\alpha_4) \frac{H}{H_c}+(\alpha_3+\alpha_4)\frac{H^2}{H_c^2}\right]\,,  \\
  p_g &=& m_g^2 M_P^2 \left[6+4\alpha_3+\alpha_4-(3+3\alpha_3+\alpha_4) \frac{H}{H_c}\left(3+\frac{{\dot H}}{H_c^2}\right)
  +(1+2\alpha_3+\alpha_4)\frac{H^2}{H_c^2}\left(3+\frac{{\dot H}}{H_c^2}\right)\right. - \nonumber \\
    & & \quad \quad \left.  -(\alpha_3+\alpha_4)\frac{H^3}{H_c^3}\left(1+\frac{{\dot H}}{H_c^2}\right)\right]\,, \nonumber
 \end{eqnarray}
where $H={\dot a}/a$ is the expansion rate of the Universe, and $H_c$ is a constant.
Cosmological solutions of the field equations and applications have been studied in  \cite{derham,volkov,langlois,gong,strauss,kobayashi,gratia,comelli,damico,lin,koyama,lambiaseBBNmassgrav}.

The modified cosmological field equations read
 \begin{equation}\label{fieldseqs}
    3 H^2=\frac{8\pi}{M_P^2} (\rho_m + \rho_g)\,, \quad 2{\dot H}+3H^2= -\frac{8\pi}{M_P^2}(p_m +\rho_g)\,.
 \end{equation}
The dot stands for the derivative with respect to $t$. The combination of the fields equations with the continuity equations yields
 \begin{equation}\label{criticalpoint}
    \left[m_g^2 \chi\left(\frac{H}{H_c}\right)-2H^2\right]\frac{{\dot H}}{H}=(1+w_m)\rho_m\,,
 \end{equation}
where $w_m=p_m/\rho_m$ and
 \begin{equation}\label{chi}
    \chi \equiv \frac{H}{H_c}\left[3+3\alpha_3+\alpha_4-2(1+2\alpha_3+\alpha_4)\frac{H}{H_c}+(\alpha_3+\alpha_4)\frac{H^2}{H_c^2}\right]\,.
 \end{equation}
An interesting consequence of the theory is that the presence of a tiny mass
of the graviton gives rise to a term that play the role of cosmological constant ($p_g \simeq -\rho_g$ \cite{volkov,langlois,gong,strauss,kobayashi,gratia,comelli,damico,lin,koyama}).
Therefore, as a {\it modified} theory of gravity, it allows to account
for the observed acceleration of the present Universe, without invoking exotic matter.

\section{Dark Matter}

As we have seen, CMB radiation physics is able to measure the spatial curvature of the Universe, the expansion rate of the Universe,
nature and spectrum of the primordial fluctuations. However, there is strong evidence that most of the mass in the
Universe is some non-luminous DM of yet unknown composition, and these cosmological and astrophysical observations are independent
of the CMB. The bulk of this dark matter is of non baryonic nature, which means that it contains no atoms and that it does not interact
with ordinary matter via electromagnetic interaction, but only gravitationally.
The nonbaryonic dark matter includes neutrinos, and may also include hypothetical entities such as axions, or supersymmetric particles.
Unlike baryonic dark matter, nonbaryonic DM does not contribute to the formation of the elements in the early Universe (BBN) and so its
presence is revealed only via its gravitational attraction. In addition, in the case in which DM is composed by supersymmetric particles,
then they can undergo to annihilation interactions with themselves resulting in observable by-products such as photons and neutrinos
(giving so an indirect detection of DM existence \cite{fornasa-bertone}).

Candidates for non-baryonic DM must satisfy several conditions: i) they should be neutral, (otherwise they would
interact electromagnetically); ii) they should not have color charge (otherwise they could form anomalous nuclear
states); iii) they should be stable on cosmological time scales (otherwise they would have decayed by now); iv) they
should interact very weakly with ordinary matter (otherwise they would not be dark), and, finally, v) they should
have a suitable relic density. If we consider the Standard Model (SM) of Particles, neutrinos seem to be the prominent
DM candidate as they interact only weakly and have an extremely low mass (the exact value of neutrinos' mass has
yet to be measured, but it is clear, from neutrino oscillation measurements, that their mass is non-zero). There are
many sources of neutrinos in the Universe, nevertheless they can account only a small fraction of DM. First of all,
they are "hot" since they move at relativistic velocities. Thus they should have had a strong effect on the instabilities
that generated the primordial cosmological objects during the earliest Universe, in the sense that galaxy clusters
would have formed before galaxies and stars. This is in contrast with most theories and measurements, which are,
instead, supported by a model based on cold DM, consisting of particles which move at non-relativistic energies. Thus
small-scale perturbations are not suppressed, allowing an early start of structure formation. Secondly, we know that
neutrino mass is rather small. The most recent upper bound on electron neutrino mass is $m_{\nu_e} < {\cal O}(1)$eV (95\% c.l.) while
the experimental limits on the muon and tau neutrino are even weaker. So neutrinos cannot explain the gravitational
effects DM is responsible for. However, extremely, high energetic neutrinos are of interest for DM search as they are
among the secondary particles created in the annihilation of other DM candidates.

The missing mass necessary to explain the observational data is present on all cosmological scales and occurs in {\it flat rotational curves} in Galaxies, {\it gravitational potential} which is responsible of confinement  of Galaxies and hot gas in cluster, {\it gravitational lenses} in clusters, {\it gravitational potential} necessary to form structure starting from tiny primeval perturbations. In what follows, we shall analyze in more details these topics which motivate the introduction of the concept of DM.

Hot dark matter cannot explain how individual galaxies formed from the Big Bang. The microwave background radiation as measured by the COBE and WMAP satellites indicates that matter has clumped on very small scales. Fast moving particles, however, cannot clump together on such small scales and, in fact, suppress the clumping of other matter. Hot dark matter, while it certainly exists in our Universe in the form of neutrinos, is therefore only part of the story.
The Concordance Model requires that, to explain structure in the Universe, it is necessary to invoke cold (non-relativistic) dark matter. Large masses, like galaxy-sized black holes can be ruled out on the basis of gravitational lensing data. However, tiny black holes are a possibility [22]. Other possibilities involving normal baryonic matter include brown dwarfs or perhaps small, dense chunks of heavy elements; such objects are known as massive compact halo objects, or "MACHOs". However, studies of big bang nucleosynthesis have convinced most scientists that baryonic matter such as MACHOs cannot be more than a small fraction of the total dark matter.

\subsection{Dark Matter and astrophysics}

In this Section we review some DM effect in astrophysical context:

\begin{itemize}
  \item {\it Dark Matter in Galaxies}

The spiral Galaxies are gravitational bound system which are stable and are composed by starts and interstellar gas. Stars and gas
move around the galactic center with an orbit almost circular, and are localized in a tiny disc. They constitute the most of the observable matter in Galaxies. Evidence of the rotation to which Galaxies undergo comes by observing the Doppler shift of (integrated) starlight and the radiation (with wavelength $\lambda=21$cm) from interstellar hydrogen gas. Within the framework of Newtonian dynamics, one finds that the radial dependence of the velocity of matter rotating in a a disk is $v=\sqrt{GM/r}$, i.e. $v(r)\sim 1/\sqrt{r}$. Stars and gas orbiting Galaxies do not follow this law. In fact, far away from the center ($\sim$ 5 kpc)  rotational curves are constant or are still rising.

The solution to this issue that attracted most attention relies on the idea that there exists a large amount of non-luminous
DM, beyond the detected stars and hydrogen could. Estimations on the luminous fraction of Galaxies is $\Omega_{lum} < 10^{-2}$ (bound obtained by radiation of baryonic matter in the visible, infrared and $X$-ray spectra), while by studying the internal dynamics of Galaxies one obtains that the latter are embedded in an extensive haloes of DM. Estimations on the halo fraction give $\Omega_{halo}> 3\times 10^{-2} - 10^{-1}$. The presence of haloes DM implies to reconsider the previous calculations leading to $v(r)$. In fact, assuming that the mass $M$ is not constant, but radially distributed with the following $r$-dependence $M(r)\sim r$, then one gets the observed constant radial velocity:
 \[
 v(r)= \sqrt{\frac{GM(r)}{r}}\simeq constant\,.
 \]
The radial density profile of the DM distribution is therefore $\rho=M/V\sim r^{-2}$. It is remarkable that this distribution is that one obtained if Galaxies were surrounded by an halo formed by an isothermal gas sphere (where the gas pressure and gravity were in thermal equilibrium). As arises from the observed rotation curves, the density plateau or core near the center is not definitively understood, and indeed there exist several profiles which fit the numerical simulations: the Navarro-Frenk-White profile \cite{white} $\rho(r)=\displaystyle{\frac{\rho_0}{x(1+x)^2}}$, the Moore {\it et al.} profile \cite{moore} $\rho(x)=\displaystyle{\frac{\rho_0}{x^{3/2}(1+x^{3/2})}}$, with $x=r/r_0$ ($r_0$ is a constant).

Of course, alternative solutions have been proposed to explain the flatness of the rotational curves of Galaxies, but they present some problems. Here we report just two example (see for example \cite{roos} for more details):
 \begin{itemize}
   \item The gravitational physics we know and apply to our solar system does not work on cosmological scales and require a deep revision. This means that or the inverse square law of gravitational force or the Newton' assumption that $G$ is a constant must be modified. However, these modifications would have to be strong at large scales, with the consequence that the cosmic shear would greatly enhanced. This picture however does not fit the observational data.
   \item The spiral Galaxies have magnetic fields, which extend till regions where the interstellar gas density is low. This condition is necessary in order that the magnetic field may modify the gas dynamics \cite{battaner}. These arguments, however, cannot work since magnetic fields affect only haloes, but not the velocity distribution of stars. Moreover, the mechanism requires a enough large strength of the magnetic fields, which have not been found yet in Galaxies.
    \end{itemize}

Evidence of DM can be found also in gravitational systems formed by a small number of Galaxies, rich clusters, and local superclusters. A discussion on these topics can be found in \cite{bertone-silk} (and reference therein) - see also \cite{roos}.

\item {\it Dark Matter and Gravitational Lensing}

The method of gravitational lensing is based on the idea that the trajectories of the light (photons) are bent by mass distribution. The latter is directly related to the deflection angle, given by $\alpha=4GM/b$, where $b$ is the impact parameter and it is much larger than the Schwarzschild radius $r_S=2GM$ of the lens. The gravitational lensing allows to determine the mass of Galaxies by measuring the distorsions of the background Galaxies
generated by the lensing. More important, the mass can be obtained without taking into account for the dynamics of the cluster under consideration.

The lensing turns out to be a powerful technique to looking for the before mentioned MACHO. This class of objects includes Jupiter-like planets, brown draft (undersized star too light to ignite thermonuclear reactions typical of nuclei stars), o dead stars (such as white dwarfs, neutron stars, black holes).
When a MACHO crosses the line-of-sight between a star (which plays the role of source) and the Earth, the MACHO acts as a gravitational (micro)lens bending the light emitted by the star. The intensity of the light undergoes to an amplification $A=\displaystyle{\frac{2+u^2}{u\sqrt{u^2+4}}}$, where $u=r/r_E$ and $r_E$ is the Einstein ring radius.

In this context, it is clear that even if the MACHOs existence are in principle possible in Galaxies, one has to monitor several millions of stars (this estimations is related to the optical depth for microlensing of the galactic halo which is $\tau \sim 10^{-6}$). Since the dark halo cannot be made up only by MACHOs, non-baryonic DM should be present out there. This kind of investigations are nowdays very active \cite{MACHO}.

\item {\it Dark Matter and  structure formation}

If one assumes that the primordial density fluctuations in a medium made by baryons are responsible for the formation of Galaxies, then the amplitude of these fluctuations must have been very large since the concentration of baryonic matter is very small. But the adiabaticy condition requires that these fluctuations must have an amplitude also very large when considered in the CMB. As a consequence, one should find an extremely large CMB anistrotopies todays. This argument therefore ruled out the Galaxy formation induced by baryonic matter.

\end{itemize}

\subsection{Dark Matter  candidates}

The most of DM candidates are, as already pointed out, of non-baryonic nature. The main
distinction is of these candidate is {\it hot DM} and {\it cold DM}. The former follows
when the constituent of DM move at relativistic speeds at the time galaxies could just
start to form. The latter, on the contrary, follows when the DM constituents move non-relativistically at that time. This distinction has important consequences for structure formation.
Experimental studies on Galaxy formation may provide an important
hint on whether DM is hot or cold. Hot DM can cluster only when it has cooled to non-relativistic speeds. N-body simulations of structure formation in a Universe dominated by hot DM, fail to
reproduce the observed structure.
The non-baryonic cold DM candidates are basically elementary
particles which have not yet been discovered. There many candidates for non-baryonic DM. The aim of this section is to recall some properties of these candidates (see the reviews \cite{bertone-silk,kamionkowski,jungman,ellis200,bergstrom2000,mavromatos2003}).

\begin{itemize}
  \item {\it Active neutrinos }

There are strong evidences of the existence of neutrinos, coming in particular from experiments on neutrino oscillations. Neutrinos, as well known, play a fundamental role at cosmological \cite{dolgov} and astrophysical \cite{raffelt-review} scales, and might represent a good candidate for DM.

For non-relativistic neutrinos, i.e. their masses are larger than the present temperature, the neutrino energy density is given by
 \[
 \rho_\nu =\sum_i m_{\nu i} n_{\nu i}\,.
  \]
(The index $i$ runs over all neutrino flavors). The neutrino energy density $\rho_\nu$ must satisfies the following constraint: it must be smaller than the mass density, $\rho_\nu < \rho_m$, which implies \cite{gershtein}
 \[
 \Omega_\nu h^2 = \sum_{i=1}^3 \frac{m_{\nu i}}{93 \,\,\mbox{eV}}
 \]
The tritium $\beta$-decay experiments provide
the best (laboratory) constraint on neutrino masses \cite{Troitsk,lambiase-cuesta}: $m_\nu < 2.05$ eV (95\% C.L.).
This upper bound holds for all three mass eigenvalues \cite{beacom,nir} (to accounting for the mass difference of solar and atmospheric neutrinos \cite{nir}), and allows to infer the upper bound on the total neutrino relic density
 \[
 \Omega_\nu h^2 < 0.076\,.
 \]
This value implies that neutrinos are simply not abundant enough to be the dominant component of dark
matter.

CMB anisotropies and large scale structures data provide a more stringent constraint on the neutrino relic density
 \[
 \Omega_\nu h^2  < 0.0067\,, \qquad \qquad  (95\% C.L.)\,.
  \]
The upper bound on neutrino mass that can be obtained is (for three degenerate neutrino species) $m_\nu <0.23$ eV.

However, if one allows for extra neutrino interactions, such as the coupling of neutrinos to a light boson, the neutrino mass limits arising from large scale structure can be evaded \cite{beacom-bell-dodelson}. Since neutrinos are relativistic collisionless particles, they erase (moving from high to low density regions) fluctuations below a scale of $\lambda_{\text{free-streaming scale}} \sim 40 \text{Mpc} (m_\nu/30\text{eV})$ \cite{bond-silk}. This would imply that big structures form first  in the Universe (a scenario termed in literature as the top-down formation history of structure). The fact that our galaxy appears to be older than the Local Group \cite{peebles-scince}, and the discrepancy between the predicted late formation of galaxies, at red-shift $z \lesssim 1$, against observations of galaxies around $z > 4$
\cite{bond-szalay}, is a further argument against neutrinos as a viable dark matter candidate.

\item {\it Sterile neutrinos }

It was suggested about 15 years ago \cite{dodelson-widrow} that sterile neutrinos could be favorite candidate of DM.
These particles do not interact weekly with other particles, as the active neutrinos, apart from mixing which could
favor their existence. It is remarkable that enlarging the Standard Model of particle physics to including sterile neutrinos and fine-tuning the parameters of neutrino sector, one is able to
explain all known facts in high energy physics and standard cosmology with Inflation
\cite{asaka2005} (see also \cite{sterile-neutrinos-altri}).
Sterile neutrinos can explain the observed velocities of pulsars \cite{kusenko}. They can also play a key role in baryogenesis \cite{akhmedov} and in the formation of the first stars \cite{neutrin-sterile-stars}. It is worth noting that unlike many other candidates for dark matter, sterile neutrinos have a nonzero free-streaming length that depends on their mass and the production history \cite{free-streaming-sterile} .
As discussed in \cite{DM-sterile,asaka2005,sterile-neutrinos-altri}, sterile neutrinos with masses in the keV range can account for cosmological dark matter (for a recent review, see \cite{kusenko2009}.

\item{\it Axions }

The axion is an {\it elusive} particle which was introduced for curing the "Strong CP problem \cite{peccei,weinberg,wilczek}. At the moment, it is one of the two best candidate particles for DM. It is expected that axions are particles that weakly interact with ordinary matter (and therefore,  in the early phases of the Universe evolution, they were in not thermal equilibrium) \cite{bradley}. Their mass is very light $m_a \lesssim 10^{-2}$eV, a bound inferred via laboratory experiments, stellar cooling, dynamics of SN1987A (see the recent review \cite{kim} and \cite{raffelt-review,bradley,raffelt-axion,rosenberg,turner1}).

Since the first propose of the possible existence of axion particles, other models based on (very) light, neutral spin zero, axion-like-particles have been proposed (see for example \cite{anselm}. Recently, analyses of starlight polarization \cite{burrage}, and the distribution and spectrum of high
energy cosmic rays \cite{fairbairn} have provided tentative evidence
for axion-like-particles. Moreover, from the astrophysical point of view, axions are considered as
a fundamental component of DM in what it provides the mechanism  for which the high-energy cosmic ray photons (the gamma rays) can safely arrive and be detected on Earth \cite{aspera}.

\item{\it Supersymmetric particles }

The Supersymmetry (SUSY) is a symmetry between fermion and bosonic particles \cite{ferrara}. It was proposed to solve the hierarchy problem of the Standard Model \cite{maiani}, i.e. why the mass of the $W$ boson is lesser than the Planck mass, $m_W \ll m_P$ , or equivalently, why $G_F \sim
1/m_W^2 \gg G_N = 1/m_P^2$? ($G_F$ is the Fermi coupling constant). An equivalent formulation relies on the question of why the Coulomb potential in an atom is so much greater than the Newton potential: $e^2 \gg G_N m^2$? ($m$ is a typical particle mass). The existence of SUSY leads to the possibility that new particles might exist, which are super-symmetric partners of particles of the Standard Model.

There are some phenomenological hints that SUSY might manifest  at
the TeV scale \cite{ellis-lectures}: the strengths of the different Standard Model interactions, as measured at LEP,
precision electroweak data prefer a relatively light Higgs boson weighing less than about 200 GeV,
the astrophysical necessity of cold DM. In the last case, indeed,
SUSY could provide a neutral, weakly-interacting massive particle, called WIMPs (weakly interacting massive particles).

More precisely, WIMPs are stable particles which arises in supersymmetric extension of the Standard Model. WIMP masses are typically in the range 10 GeV - 10 TeV, and they have
interactions with ordinary matter which are characteristic of the weak interactions.
The most promising WIMP candidate is the neutralino. Below a short list of supersymmetric particles
candidate to DM (see also \cite{steffen} and references therein):

\begin{itemize}
  \item {\it Axinos} -  This particle is the fermionic (spin-1/2) superpartner of the axion (the bosonic  superpartner is called saxion - axino and saxion particles are bundled up in a chiral superfield). Depending on the model and the SUSY breaking scheme, the axino
     mass is in the range  eV $\lesssim m_{\text{axino}}\lesssim $ GeV \cite{axinomass}
     Axinos as warm and hot DM candidate has been discussed in \cite{bonometto-axion}. The properties of cold axino DM as a possible candidate (provided that the reheating temperature is quite low) can be found in \cite{covi}. See also \cite{kim,bertone-silk} and reference therein.
  \item {\it Gravitinos} - In local supersymmetic models (i.e. passing from a global
   to a local symmetry which leads to supergravity), gravitinos are the superpartners (spin-3/2)
  of the gravitons.  The mass of gravitino (strongly depending on the SUSY breaking
    scheme) ranges from $\sim$ eV scales to $\gtrsim $ TeV scales \cite{gravitinomass}.
    However, since they interact only gravitationally, it is very difficult to observe \cite{feng}.

  One of the big problem of (long lived-)gravitinos is that they can be in disagreement with the predictions of the Big Bang Nucleosynthesis, because in some scenarios they can destroy abundances of primordial light elements \cite{gravitinos-BBN}. Another problem is the overproduced of gravitinos in the early Universe if the temperature of the reheating epoch is not sufficiently low \cite{gravitino-reheating}. However, it must pointed out that some models can avoid these problems \cite{gravitino-avoidproblem}.
    \item {\it Neutralinos} - Neutralinos are the most promising and suitable candidate to DM \cite{steffen,bertone-silk,neutralino-candidate}.
    The {\it lightest} neutralino appears in the so called minimal
supersymmetric Standard Model (MSSM) as the
lightest mass eigenstate among the four neutralinos being
mixtures of susy partners: the bino, the wino, and the neutral
higgsinos.  It is a fermion (spin-1/2) which interacts only weakly with other particles, with a mass which depends on the gaugino mass parameters, on the ratio of the two MSSM Higgs doublet vacuum expectation values, and the higgsino mass parameter. Estimations give
$m_{neutralino}= O(100$ GeV), and therefore it is classified as WIMP\footnote{Actually, two conditions must be satisfied: 1) the neutralino must be the lightest supersymmetric particle
a condition that occurs in a broad region of the space
parameter of minimal supergravity model; 2)
the R-parity must be conserved, which means that the neutralino must be a stable particle.
The R-parity takes the values +1 for all conventional particles and -1 for all sparticles \cite{fayet}. The conservation of R parity can be related to that of baryon/lepton numbers $B/L$, since
$ R = (-1)^{3B+L+2S}$, where $S$ is the spin. There are three important consequences of R conservation \cite{ellis-lectures}:
 1. Sparticles are always produced in pairs;
 2. Heavier sparticles decay to lighter ones;
 3. The lightest sparticle (LSP) is stable, because it has no legal decay mode.
Neutralino exists in thermal equilibrium and in abundance in the
early Universe, when the temperature of the Universe exceeds the mass
of the particle.}.

  \item {\it Sneutrinos} - In the Minimal Supersymmetric Standard Model, these particles are the superpartners of the neutrino of the Standard Model. The (left-handed) sneutrinos are not a viable dark matter candidate essentially for two reasons
related to their sizable coupling to boson vector $Z$: in the first case, the strong coupling  induces a too rapid annihilation that generates a very small relic abundance, whereas in the second case, the coupling gives rise to a large scattering cross section of nucleons that
are excluded by direct DM searches \cite{falk} (the detection cross section can be reduced if one accounts for a lepton number violating operator, as shown in \cite{hall}).
However, there has been a renew interest and strong motivation to re-consider
an extension of the Minimal Supersymmetric Standard Model owing to neutrino oscillations experiments which showed that neutrino are massive. These models can be
obtained by introducing right-handed neutrino superfields. In this context,
several models have been proposed to re-consider sneutrino DM by reducing its coupling with the Z boson (see for example \cite{sneutrinos}).

\end{itemize}

As a conclusion of the possible candidates for DM, we recall other possibilities: Light scalar DM \cite{355}, little Higgs model (which represents an alternative mechanism to SUSY to stabilize the weak scale) \cite{38}, Superheavy DM or wimpzilla \cite{144}, Q-balls \cite{350}, mirror particles \cite{237}, CHAMPs (charged massive particles) \cite{182}, self-interacting DM \cite{173}, D-matter (particle-like states originating from D branes) \cite{449}, cryptons (bound states in the hidden sector of superstring-model) \cite{212}, superweakly interacting DM \cite{234}, brane world DM \cite{141}, four generation of neutrinos \cite{321}.

\item{Dark Matter in Brane Cosmology}

In is worth to also quote the correlation between DM and the {\it warped extra dimensions} model.
It was proposed in Ref. \cite{randall} in order to explain the large hierarchy between
the electroweak scale and the Planck scale\footnote{More specifically, in the warped extra dimensions scenario, it is assumed the existence
of an extra dimension compactified on a $S_11/Z_22$ orbifold,
with two branes sitting on each orbifold fixed point. The brane at
$y = 0$ is called the Planck brane, while the brane at $y = \pi r_c$ is called the TeV
or SM brane. With an appropriate tuning for cosmological constants in the bulk
and on the branes, we obtain the warped metric
 \[
ds^2 = e^{-2 \kappa |y|}\eta_{\mu\nu} dx^\mu dx^\nu -dy^2\,,\qquad \eta = diag(-1, —1, —1, —1)
 \]
This type of geometry is called non-factorizable because the metric of the 4D subspace is y-dependent.
In the simplest version of the RS model it is assumed that the fields of the Standard Model live on
the TeV brane, while gravity (gravitational fields) lives everywhere.}. The idea of this model is that matter is, by assumption, localized in
a brane embedded in the bulk (a space with larger with respect to brane).
In the last years there has been a growing interest in studying the non-conventional brane cosmology on the relic
abundance of dark matter \cite{okada}. An interesting consequence of the non-conventional brane cosmology is that
the evolution of the Universe is quite different from the standard one (i.e. FRW cosmology) based on Einstein equations.
In fact, the Friedman equation of a brane embedded in five dimensional
(5D) warped geometry is  \cite{allbranecosmology}
 \[
 H^2=\frac{8\pi G_4}{3}\rho \left(1+\frac{\rho}{\rho_0}\right)-\frac{k}{a^2}+\frac{C}{a^4}\,.
 \]
where  $\rho$ is the energy density of {\it ordinary matter} on the brane, $\rho_0$ is the brane tension, $G_4$ is
the 4D Newton coupling constant, $k$ the curvature of the three
spatial dimensional, and finally $C$ is a constant of integration which is called {\it dark-radiation} (constrained by nucleosynthesis analysis).
At a high energy regime $\rho\gg \rho_0$ one gets  $H \sim \rho$, so that the evolution of the scale factor is strongly modified in these models, leading to a dark matter relic abundance, that may occur during the early phases of
the Universe.

\end{itemize}


\section{Conclusions}

The recent observations as well as the huge amount of cosmological data have given rise
to the problem of DE and DM, Without any doubt they represents an intriguing puzzle of modern cosmology and particle physics, and have consequently stimulated the search of models able to shed new light on the effective picture of the Universe.

As we have seen, type Ia Supernovae, anisotropies in the cosmic microwave background radiation,
and matter power spectra inferred from large galaxy surveys represent the strongest
evidences for a radical revision of the cosmological standard model. In particular, the
concordance $\Lambda$CDM model predicts that baryons contribute only $4\%$ of the total matter
-energy budget, while the exotic cold dark matter represents the bulk of the matter
content ($25\%$) and the cosmological constant  plays the role of the so called dark
energy ($70\%$).

Several theories and models have been proposed in literature to solve, or at least to try to solve, the puzzle of the nature of DE and DM, the former ranging from scalar fields playing the role of time-dependent cosmological constant to modification of gravity sector, the latter invoking supersymemtric and exotic particles created in the early phases of the Universe evolution.
All these scenarios, although based on deeply different physics, are able to account for the available astrophysical and cosmological data. Despite this, however, these models have been unable till now  to give a definitive answer or solution, and much work is necessary in the next future to understanding these fundamental issues of our Universe.



\end{document}